\documentclass[usegraphicx]{mn2e}
	
\def\hii{{\sc H\thinspace ii}}
\def\hi{{\sc H\thinspace i}}
\newcommand\rmin{R_{\rm min}}
\newcommand\rmax{R_{\rm max}}

\newcommand\SFR{{\rm SFR }}

\def\simless{\mathbin{\lower 3pt\hbox
   {$\rlap{\raise 5pt\hbox{$\char'074$}}\mathchar"7218$}}}   
\def\simgreat{\mathbin{\lower 3pt\hbox
   {$\rlap{\raise 5pt\hbox{$\char'076$}}\mathchar"7218$}}}   
\def\etal{{\rm et al.}}

  \makeatletter
  \ifx\CUP@mtlplain@loaded\undefined  
  \makeatother  
    
  \else
    
  \fi

\loadboldmathitalic 
\loadboldgreek      
  
\makeatletter
\ifx\CUP@mtlplain@loaded\undefined  
  \newfont\bit{cmbxti10 at 9pt}
\else
  \newfont\bit{mtbxti10 at 9pt}
\fi
\makeatother  
 
\title[Star formation threshold for radiative feedback] 
{Galactic porosity and a star formation threshold
	for the escape of ionising radiation from galaxies}
  \author[C.J. Clarke \& M.S. Oey]  {Cathie Clarke$^1$ and M. S. Oey$^2$ \\
$^1$Institute of Astronomy, Madingley Road, Cambridge CB3 0HA\\
$^2$Lowell Observatory, 1400 W. Mars Hill Rd., Flagstaff, AZ, 86001, USA}
\date{19 August 2002; Accepted to MNRAS}
\pagerange{\pageref{firstpage}--\pageref{lastpage}}
\pubyear{    }

\def\LaTeX{L\kern-.36em\raise.3ex\hbox{a}\kern-.15em
    T\kern-.1667em\lower.7ex\hbox{E}\kern-.125emX}

\let\leqslant=\leq

\begin{document}

\label{firstpage}

\maketitle

\begin{abstract}
The spatial distribution of star formation within  galaxies
strongly affects the resulting feedback processes.
Previous work has considered the case of a single, concentrated
nuclear starburst, and also that of distributed single supernovae (SNe).
Here, we consider ISM structuring by SNe originating
in spatially distributed clusters having a cluster membership spectrum
given by the observed \hii\ region luminosity function.  We show that in
this case, the volume of \hi\ cleared per SN is considerably greater
than in either of the two cases considered hitherto.

 We derive a simple relationship between the ``porosity'' of the ISM
and the star formation rate (SFR), and deduce a critical $\SFR_{\rm crit}$, at
which the ISM porosity is unity.  This critical value describes
the case in which the SN mechanical energy output over a timescale $t_e$
is comparable with the ISM ``thermal'' energy contained in random
motions; $t_e$ is the duration of SN mechanical input per superbubble.
This condition also defines a critical gas consumption timescale $t_{\rm exh}$,
which for a Salpeter IMF  and random velocities of
$\simeq 10\ {\rm km\ s}^{-1}$ is roughly $10^{10}$ years. 

 We draw a link between porosity and the escape of ionising radiation
from galaxies, arguing that high escape fractions are expected
if $\SFR \simgreat \SFR_{\rm crit}$.  The Lyman Break Galaxies,
which are presumably subject to infall on a timescale $<t_{\rm exh}$, 
meet this criterion, as is consistent with the significant leakage of
ionising photons inferred in these systems.  We suggest the utility of
this simple parameterisation of escape fraction in terms of SFR
for semi-empirical models of galaxy formation and evolution and for
modeling mechanical and chemical
feedback effects.

\end{abstract}


\section{Introduction}
 
Feedback from supernovae (SNe) is a major ingredient of most contemporary
models for galaxy formation and evolution.  One of the main motives
for its inclusion is the need to reconcile the predictions of CDM
cosmology with the observed galaxy luminosity function (e.g. Cole
et al 1994, 2000; Efstathiou 2000): the overproduction of dwarf
galaxies in CDM models can be alleviated if star formation is inefficient
in low mass systems, and the explosive energy input from SNe 
provides an obvious mechanism
for ejecting gas from the shallow potentials of dwarf galaxies. SNe
also return metals to the ISM, and if this process is coupled
with the strong outflows posited above, can also provide a mechanism
for enriching the IGM (e.g., Dekel \& Silk 1986; Madau, Ferrara and Rees 2001).

  Another aspect of SN-driven feedback that has received 
rather less attention is its  possible relation to the escape of
ionising radiation from star forming galaxies. It is currently
unclear whether stellar sources or quasars are mainly responsible
either for the re-ionisation of the Universe at high redshift
or for the present day ultraviolet background (Madau, Haardt and
Rees 1999), the answer depending critically on the assumed
fraction of Lyman continuum photons that are able to escape
star forming galaxies (Giallongo et al 1997; 
Madau and Shull 1996; Bianchi et al 2001).  
The simplest parameterisation
of this problem involves assuming a constant escape fraction for all
galaxies, but there are obvious reasons for supposing that in
reality the escape fraction should depend on galactic parameters.
In particular, the escape fraction, which depends on the distribution
of neutral hydrogen along the line of sight, is likely to be strongly
affected by the the re-structuring of the ISM effected by SN
explosions. At a qualitative level, it would seem evident that higher
escape fractions should be expected in systems with a higher star
formation rate, since the disintegration of interacting SN-driven
bubbles can in principle  
open up lines of sight through which Lyman continuum photons can 
leak.  Some observational support for the notion is provided by the
recent detection of Lyman continuum emission in the composite
spectra of Lyman Break Galaxies (Steidel et al 2001), which include
systems undergoing vigorous star formation.  

To date, however, calculations of escape fractions in star forming
galaxies do not exhibit a strong  dependence of escape fraction on
star formation rate (SFR); in the case of photoionisation calculations in
a smoothly stratified ISM (Dove and Shull 1994; Ricotti and Shull 2000;
Wood and Loeb 2000) the derived low escape fractions 
are not strongly dependent on the SFR, provided this
exceeds the threshold value at which the resulting \hii\ regions cease to
be ionisation bounded. Recent calculations by Ciardi et al (2001),
demonstrate an even weaker dependence of escape fraction on SFR
in the case of an inhomogeneous, fractal ISM. In all these
calculations, however, the assumed ISM
structure is independent of star formation activity. The effects of
mechanical feedback on the escape of ionising radiation have so far
only been considered in the case of a coeval burst of star formation,
where the source of ionising radiation and SN energy originate
in the same region (Tenorio-Tagle et al 1999; Dove, Shull and Ferrara 1999): 
here the temporary trapping of the ionisation
front in the wall of the SN-driven bubble and the steep
temporal decline of the ionising radiation from the burst combine
to also produce a low escape fraction.

  In this paper, we consider the case  where the photons
from each  OB association impact on an ISM that has been structured
by SN-blown superbubbles, reflecting its star formation history
over the last  $10^7-10^8$ years. We thus draw a link between the
{\it{porosity}} of the ISM, i.e., the fraction of the ISM that is
devoid of \hi\ due to the expansion of SN-driven bubbles, and
the resulting escape fraction of ionising radiation. In order to
quantify this link in more detail, one requires both  3D  hydrodynamic
calculations, which model the break up of bubble walls following the
interaction of adjoining bubbles, and radiative transfer calculations,
which calculate the escape of ionising radiation from the resulting
distribution of HI (see Fujita et al 2002).  Here, we attempt neither 
of these exercises but instead, having posited that the escape fraction
should rise steeply as the porosity of the ISM approaches unity, set up
a model of the ISM which allows the porosity to be readily calculated
as a function of SFR and galactic parameters.

In this work, we focus on
the {\it volume,} $V_{\rm sn}$ of the
ISM that is cleared of neutral hydrogen per unit SN explosion.
This value depends strongly on the spatial distribution and clustering
properties of the SN progenitors.  Since superbubbles expand to
the point that 
they are in rough pressure equilibrium with the ISM, it follows that an
upper limit to this quantity is $V_{\rm max}$, the volume of ISM
originally containing a thermal energy equal to the SN energy
($\sim 10^{51}$ ergs).  Note that in this work, we consider ``thermal
energy'' to refer to both macroscopic and microscopic random motions
in the ISM. 

To date, most previous analyses have made either 
one of two extreme assumptions regarding the distribution of SN
progenitors: either they are distributed singly throughout
the galaxy  (Larson 1974; McKee and Ostriker 1977; Dekel and Silk 1986;
Efstathiou 2000) or else, as in  more recent studies (De Young \&
Heckman 1994; Suchkov et al 1994; Mac Low and Ferrara 1999;
Strickland \& Stevens 2000) they are concentrated in a single coeval,
cospatial burst of star formation. 
Each of these assumptions however predicts a volume cleared per SN
that is much less than $V_{\rm max}$, though for different reasons in the
two cases. In the case of single SNe, the limiting factor is
the cooling of the shocked ISM (Cox 1972; Chevalier 1974; Goodwin et al
2002), following which the bubble expands as a pressure driven snow plough. 
Due to these radiative losses, the final bubble encloses a volume of
ISM that is a few per cent of $V_{\rm max}$ (Cioffi et al 1988). In the
case of clustered SN progenitors, the effect 
of multiple SNe may be modeled as a continuous input of mechanical
energy, analogous to a stellar wind (McCray \& Kafatos 1987; Mac Low
\& McCray 1988), and in this case it is found that cooling is of
marginal importance.  However, in the case
of a luminous starburst in a disc galaxy, the superbubble
blows out when its size is between one and two disc scale heights. The
bubble interior is consequently de-pressured by the loss of hot gas normal
to the disc plane, and again the resulting cavity in the ISM implies
a volume cleared per SN that is much less than $V_{\rm max}$. 

    In this paper we examine the case in which SNe are distributed
according to the observed distribution of OB stars in galaxies.
Specifically, we consider spatially distributed OB associations and
superclusters whose membership numbers are inferred from the observed
luminosity function of \hii\ regions and OB associations  (Oey and Clarke
1998; McKee and Williams 1997).  The  OB association 
membership function is such that the number of associations
having numbers of stars between $N_*$ and $N_* + dN_*$ is
$ \propto N_*^{-2}$. This distribution  is similar in functional form to  
the observed mass distribution of clumps within
molecular clouds and the membership number function 
for all types of stellar clusters (Blitz 1991;
Elmegreen and Clemens 1985; Harris and Pudritz 1994; Elmegreen \&
Efremov 1997; Meurer et al. 1995).  In a previous paper (Oey and
Clarke 1997), we quantified the distribution of \hi\ hole sizes
predicted by such a 
model and found it to be in excellent agreement with the observed
distribution of \hi\ holes in the Magellanic Clouds (Oey \& Clarke 1997;
Kim et al. 1999).
This empirical success encourages
us to assume that this scale-free distribution
of OB association richness is a universal characteristic of the
ISM on all scales and in all environments. 
With such a prescription we can compute the fraction of the ISM that
is cleared of \hi\ for a given SFR (Oey et al. 2001), and also follow
the evolution of this quantity during an episode of star formation.
As we show in Section 2.1, this model implies that the volume of ISM
swept up per SN is considerably larger than
in either of the two limits that have been considered to date,
and may be a significant fraction of $V_{\rm max}$. 

Our analysis expands on the premise introduced by Oey et al. (2001),
of a critical $\SFR_{\rm crit}$, such that lower rates produce a volume filling
factor of holes that is considerably less than unity.  Under these
circumstances, we surmise that the escape fraction of ionising
radiation will be low, as found in photoionisation  
calculations based on a smoothly stratified ISM.  If the star formation
rate exceeds this value, the ISM becomes filled with hot bubbles
and we speculate that the escape fraction from the ISM may rise considerably
at this point, as the widespread merging of bubbles leads to the break up
of shell walls through a variety of instabilities.
Disc systems may be able to continue to produce stars at such
a rate, but in spheroidal systems, such as molecular clouds or proto-globular
clusters, this seems unlikely, as the energy input into the gas 
from SNe at this point is comparable with the
self-gravitational binding energy of the system.  We investigate, in a
crude analysis, how the finite time required to clear the star forming
region of \hi\ affects the number of ionising photons that can escape
from the system, taking into account the rapid temporal decline of
the ionising luminosity produced by a stellar population.  
Our approach is thus especially useful in providing estimates of
escape fractions and star formation efficiencies for regions that may
be below the resolution limit for numerical simulations of galaxies.

The structure of the paper is as follows.
In Section 2 we set out the model for the growth of superbubbles and
derive an expression for the porosity of the galaxy as a function of
time, SFR, and ISM parameters, including an analysis
of how the results are modified in disc systems, and establish
the existence for a {\it critical star formation threshold}. 
In Section 3 we assess the consequences of the model for the star formation
efficiency and escape of ionising photons from galaxies.
In Section 4 we apply the results of the foregoing sections
to a variety of star forming regions, including Lyman break
galaxies, starbursts and giant molecular clouds. Section 5
summarises our conclusions.

\section {Porosity related to star formation:  an analytical approach}

Oey \& Clarke (1997; hereafter OC97) estimate galactic porosity as the
volume of superbubbles generated by OB associations relative to
the simple geometric volume of the host galaxy.  Here, we examine the
porosity in more detail and relate it specifically to the ISM
thermal energy and mass. 

\subsection{The volume of hot gas generated by star formation: steady state}

We take the mechanical luminosity function (MLF) for the SN
energies of the OB associations to be,
\begin{equation}\label{MLF}
\phi(L) = A L^{-2} \quad ,
\end{equation}
where $\phi(L)$ is the fraction of clusters with mechanical luminosity
in the range $L$ to $L+dL$.  The power-law slope of --2 is empirically
well-determined from the \hii\ region luminosity function
and stellar cluster mass function, as described above.
Such a distribution implies that equal decades in cluster
luminosity contribute equally to the total integrated mechanical
luminosity, or, equivalently, SFR, of the galaxy.  Thus the total
SFR is dominated neither by very populous nor very
sparse clusters.  We furthermore assume that this luminosity function
extends over a range of luminosities corresponding to a supernova membership
number in the range $N_{\rm min}$ ($\geq 1$) to $N_{\rm max}$.  

Following OC97, the lifetime of all superbubbles is assumed to be
roughly equal to the main-sequence lifetime of the lowest mass SN
progenitor, $t_e \sim 40$ Myr for Population I stars.  Therefore, if constant
star-forming conditions are sustained for periods in excess of $t_e$,
the differential superbubble size distribution attains a steady state,
as derived by OC97: 
\begin{equation}\label{size_s}
N(R) = A\psi L_e^{-1} R_e^{-1}\ \Bigl(\frac{R}{R_e}\Bigr)^{-3}\
	\Bigl[2(t_e + t_s) - \frac{3}{4}\ t_e\ \Bigl(\frac{R}{R_e}\Bigr)\Bigr]
	\ ,\ R \leq R_e
\end{equation}
and
\begin{equation}\label{size_g}
N(R) = 5A\psi\ L_e^{-1} R_e^{-1}\ \Bigl(\frac{R}{R_e}\Bigr)^{-6}\
	\Bigl[\frac{t_e}{4} + t_s\Bigr] \ , \ R > R_e
\end{equation}
\noindent where $\psi$ is the creation rate of the superbubbles,
$L_e$ is the luminosity of a bubble that comes into
pressure equilibrium with the ambient medium after time $t_e$
and $R_e$ is the corresponding radius of such a bubble at that
time.  It is assumed that the growth of bubbles with $L < L_e$ stall by
pressure confinement at the point that they come into
pressure balance with the ambient medium.  
After $t_e$, the SN energy stops, and the object is presumed to
survive at constant radius for another increment of time $t_s$.

For this steady-state size distribution and constant MLF, the total volume
of superbubbles depends only on $\psi$, or equivalently SFR, and
the interstellar conditions that determine $R_e$.  
The total volume of the superbubbles in a steady state is  
\begin{equation}
V_{\rm tot,ss} = \int_{\rmin}^{\rmax} \frac{4}{3}\pi R^3\ N(R)\ dR \quad ,
\end{equation}

\noindent where $R_{\rm min}$ is the stall radius of a bubble containing $N_{\rm min}$
supernovae and $R_{\rm max}$ is the size of a bubble containing
 $N_{\rm max}$ supernovae at time $t_e$.
Thus integrating equations~\ref{size_s} and \ref{size_g} gives a total
volume, 
\begin{equation}
V_{\rm tot,ss} \simeq 3\pi A\psi L_e^{-1} R_e^3\ (t_e + 2t_s) 
   \quad .
\end{equation}
\noindent Equation~\ref{MLF} is a probability distribution, so its integral is
unity, and therefore $A \simeq L_{\rm min}$, yielding,

\begin{equation}\label{Vtot}
V_{\rm tot,ss} \simeq 3\pi N_{\rm tot} \Bigl(\frac{L_{\rm min}}{L_e}\Bigr)\ 
	R_e^3 \quad ,
\end{equation}
\noindent where $N_{\rm tot}$ is the total number of superbubbles in the
steady state.

  The total number of supernovae contained in this population
of bubbles is (from equation~\ref{MLF})

\begin{equation}\label{Nsn}
N_{\rm sn} \simeq  N_{\rm tot} N_{\rm min} {\rm ln} \Bigl(\frac{N_{\rm max}}{N_{\rm min}}\Bigr)\
         \quad ,
\end{equation}

\noindent so that the {\it mean volume of ISM cleared per supernova}, $V_{\rm sn}$ is

\begin{equation}\label{Vsn}
V_{\rm sn} \simeq  \frac{3 \pi}{{\rm ln} \Bigl(\frac{N_{\rm max}}{N_{\rm 
min}}\Bigr)} \frac{R_e^3}{N_e}
         \quad ,
\end{equation}

\noindent where we have used the fact that $ L \propto N$, and we
define $N_e$ as the 
number of SNe corresponding to mechanical luminosity $L_e$.  We
will use these expressions for SN-cleared volume in deriving
the interstellar porosity below.

Equation~\ref{Vsn} shows that $R_e$ dominates $V_{\rm sn}$.  Note that
since $R_e$ is the radius at which a bubble containing $N_e$ supernovae 
comes into pressure balance with the ISM, one can roughly equate the thermal
energy of the ISM contained within $R_e$ with the total energy input 
from $N_e$ supernovae:

\begin{equation}\label{therm}
 N_e E_{\rm sn} \simeq \frac{4\pi}{3} R_e^3 u
         \quad ,
\end{equation}

\noindent where $E_{\rm sn} \sim 10^{51}$ erg is
the SN energy and $u$ is the thermal energy density in the 
ISM\footnote{Note that throughout this paper we use the term
`thermal' energy to denote energy in random
motions in the ISM, whether this is dominated by bulk cloud motions
or by motions at a molecular (thermal) level.} .
We thus deduce that $V_{\rm sn}$ is within a factor of order unity
of $V_{\rm max}$:

\begin{equation}\label{vmax}
V_{\rm max} \sim  {{E_{\rm sn}}\over{u }} \quad .
\end{equation}

\noindent which, as discussed in \S 1, is the mean  volume per supernova
for the adiabatic evolution of individual SNe.

 The finding that $V_{\rm sn} \simeq V_{\rm max}$ contrasts
with the two scenarios considered by previous authors, namely,
either distributed individual SNe, or else all SNe concentrated
in a single bubble.  The difference may readily be traced to the
fact that when one considers a realistic spectrum of cluster
richness (i.e. a MLF) there is an important volumetric contribution
from bubbles with $L \simeq L_e$. Such bubbles, which stall after
a time $t_e$, remain in the adiabatic expansion phase over their
entire SN-producing lifetime and thus represent optimal
coupling between the supernova energy and clearing of the ISM.
Note also that $V_{\rm sn}$ is
insensitive to any upper cut-off in the richness of OB associations,
provided that the MLF extends well beyond $L_e$, since the
volumetric contribution of bubbles with $L\gg L_e$ is small 
(equation~\ref{size_g}).

\subsection{Non-steady star formation}

 The above analysis may readily be modified to model an episode of
star formation that proceeds at constant rate over a timescale
$t <t_e$.  Such a situation is only relevant to star forming
systems that can switch on their star formation on timescales
$\ll t_e$ and thus applies to compact systems with short dynamical
timescales, such as molecular clouds and globular clusters, as
discussed in \S 4.1.

At time $t (<t_e)$, the transition from the size distribution given by
equation~\ref{size_s} to that of equation~\ref{size_g}
occurs at radius $R_t$ instead of $R_e$, where $R_t$
is the radius of a bubble that just stalls at time $t$.
Consequently, the total volume of superbubbles after time
$t$ is given by:  

\begin{equation}\label{vtt}
V_{\rm tot,ns} (t) = 3 \pi N_{\rm tot}(t) R_{\rm min}^2 R_t
         \quad .
\end{equation}

\noindent $N_{\rm tot}(t)$,  the total number of superbubbles produced in time
$t$, is proportional to $t$, whereas one may readily show (OC97)
that $R_t \propto t$. Consequently, the volume of hot gas created
varies {\it quadratically} with time, so that one may write

\begin{equation}\label{vtt2}
V_{\rm tot,ns} (t) = V_{\rm tot,ss} \Bigl(\frac{t}{t_e}\Bigr)^2 
         \quad .
\end{equation}

\subsection {Effect of cooling and finite scale of ISM}

  The above analysis is appropriate to an infinite ISM where the
bubble evolution remains adiabatic until it comes into pressure
equilibrium with the ISM.

   We verify the approximate validity of the adiabatic assumption  
by comparison of  the bubble stall radius with the cooling radius
given by Mac Low and McCray (1988).  The former may be written
(e.g. see OC97) in the form:
\begin{equation}
R_f =  300\ {\rm pc}\ L_{38}^{1/2} P_{\rm MW}^{-3/4} n_{\rm MW}^{1/4} \quad ,
\end{equation}
where $L_{38}$ is the luminosity normalised to $10^{38}$ erg s$^{-1}$
(equivalent to an OB association with $\sim100$ SN
progenitors), $P_{\rm MW}$ and $n_{\rm MW}$ are respectively the
pressure and number density 
of the ISM normalised to their values in the Milky Way 
($3 \times 10^{-12}$ dyne cm$^{-2}$ and $0.5$ cm$^{-3}$).
The corresponding expression
for the cooling radius of freely expanding bubbles,
using a cooling function appropriate to solar
metallicity gas, is (Mac Low and McCray 1988):
\begin{equation}
R_c \simeq 540 \ {\rm pc}\ L_{38}^{4/11} n_{\rm MW}^{-7/11} \quad .
\end{equation}
 
 These expressions imply that even in the case of
solar metallicity gas,  cooling is of marginal importance
before the point is reached where the counter-pressure of the
ISM is significant in slowing the expansion; for lower metallicity
systems, cooling would be of still less importance. 
This hence justifies  our treatment of the evolution  prior
to stalling as
approximately adiabatic. 

   We now consider how the above analysis is modified when one takes
into account the finite extent of the ISM. In general, bubbles evolve
as described above, provided their sizes remain less than the density
scale length
of the ISM. Once they grow to larger radii, their evolution is modified
by the ambient density gradient: 
in  particular,  a decreasing gradient
causes the bubble expansion to accelerate due to the decreasing inertia
of the newly swept up material. In disc-like density distributions,
for example, a  number of authors have performed 
hydrodynamical simulations of bubbles that demonstrate that bubbles
`break out' of the disc when they grow to a height between one and two
disc scale heights (e.g., Mac Low and McCray 1988). At this
point, the contents of the hot bubble interior are vented 
normal to the disc plane. Thereafter, the bubble evolution in the
disc plane is no longer adiabatically driven, but evolves in a
momentum-conserving fashion.  We now calculate
how the volume of ISM occupied by bubbles is reduced if
one takes this into account.   

 It is convenient to divide the bubble population into low luminosity
objects having $ L< L_H$, which stall at sizes less
than the disk scale height $H$, and higher luminosity objects that
break out of the disc. 
>From integration of equation~\ref{size_s} it can be seen that the total
volume of bubbles contained in objects with radius less than $R$ is
roughly proportional to $R$, and the total volume contained in bubbles that
never break out of the disc is a factor $\simeq H/R_e $ times the
total volume of bubbles that would be created in an infinite medium for
a given SFR.  

We now estimate the total volume swept out by bubbles with
$L > L_H$. Such bubbles evolve adiabatically prior to breakout
and hence the kinetic energy of the bubbles walls is proportional
to the number of supernovae that have gone off at that point. 
Since bubbles attain a fixed size scale on a timescale
$t_H$ that scales as $L^{-1/3}$\ (OC97), the  kinetic energy of bubbles 
at breakout scales as $L \times t_H \propto L^{2/3}$.  All bubbles with
$L > L_H$ break out when the volume of ISM swept up is $\sim H^3$,
so that the mass of ISM swept up at breakout is independent of 
$L$. Hence the {\it momentum} of bubbles at breakout scales simply as the
square root of the energy, i.e. as $L^{1/3}$.  Thereafter,
the bubbles evolve in an approximately momentum-conserving
fashion and then stall when their expansion velocities become of order 
the thermal speed in the ISM.  Thus, it follows that the final volume of
the bubble is proportional to the momentum at
breakout, and hence also scales as $L^{1/3}$.  We can obtain the
normalisation by noting that objects that just stall at size scale
$H$ are by definition not going to undergo further momentum-conserving
expansion, because their velocity has already declined to thermal values.
 
Thus we find that the final volume of a bubble of size $L > L_H$ is given by 
$H^3 (L/L_H)^{1/3}$ (see also Koo \& McKee 1992), 
as compared with the final bubble volume in an infinite
medium which can be written as $H^3 (L/L_H)^{3/2}$ (OC97).
By integrating each of these expressions over the MLF
(equation~\ref{MLF}), we find that bubbles that have broken out contribute
a total volume that is a factor $\simeq H/R_{\rm e}$ times
the total volume filled in the case of an infinite medium.
[Note that this analysis does not account for any continued driving by
remaining SNe, which, following breakout from the disk, could
contribute power following a momentum-conserving shell evolution
(Steigman {\etal} 1975).  
It can be shown, using the corresponding
relations from OC97, that in this case the bubbles with
$L>L_H$ then  contribute a volume fraction that exceeds the
above estimate by only a logarithmic factor ($\ln\frac{R_{\rm max}}{H}$).
The effect of continued SN driving is thus
not expected to be large, and we therefore do not consider it further
for the purpose of the rough estimates considered here.]

Thus adding together
the total contributions from bubbles that do and do not break out, 
and taking $R_{\rm max} \simeq R_e$, we
find that the volume of bubbles produced is reduced by a factor
\begin{equation}\label{fd}
 $$ f_d \sim 2 H/R_e $$
\end{equation}
In forthcoming sections, we shall apply this correction factor where
necessary in order to reduce the volume of bubbles produced per unit
SFR in disc galaxies.

\subsection{Calculation of galactic porosity}

  In order to compute the porosity of the ISM, it is necessary
to divide the volume of hot gas produced by  star formation
by the effective volume of the star forming system, $V$. Thus from
equations~\ref{Vtot} and \ref{therm}, the steady
state porosity can  be written

\begin{equation}\label{qss}
 Q_{\rm ss} \simeq \frac {f_d V_{\rm tot,ss}}{V} \simeq \frac{9}{4}\ \frac{f_d N_{\rm tot} N_{\rm min} E_{\rm sn}}{uV}
         \quad ,
\end{equation}

\noindent where $f_d$ is the factor (equation~\ref{fd}) that takes rough account of the
reduction in galactic porosity in the case of disc systems. If the mean mass
of stars produced per bubble is $m_*$, then $N_{tot}$ is related to the
star formation rate by:

\begin{equation}\label{Ntot}
N_{\rm tot} = \frac{\SFR\ t_e}{m_*}
         \quad ,
\end{equation}

\noindent whilst the product $uV$ is, by definition, the total thermal energy 
contained in the ISM of the system, $E_{\rm ISM}$: 

\begin{equation}\label{EISM}
 uV = E_{\rm ISM} = \frac{1}{2} M_{\rm ISM}\ \tilde v^2 
         \quad ,
\end{equation}

\noindent where $M_{\rm ISM}$ is the total mass in the ISM and $\tilde v$ is the
`thermal' velocity dispersion. Thus ~\ref{qss} becomes

\begin{equation}\label{qss2}
 Q_{\rm ss} \simeq \frac{9}{2}\ \frac{f_d N_{\rm min} {\rm SFR}\ t_e E_{\rm sn}}{m_* M_{\rm ISM}\ \tilde v ^2} 
         \quad .
\end{equation}

For a Salpeter IMF and $N_{\rm min} =1 $, the mean mass of stars per
bubble can be written $m_* = 150\ {\rm ln}(N_{\rm max})\ M_\odot$ and
is thus weakly (logarithmically) 
sensitive to any upper cutoff in the MLF. Here we adopt $N_{\rm max} \sim
7000$, which corresponds to the largest OB associations in the Milky
Way (McKee and Williams 1997), and which is, incidentally
about twice  $N_e$ for  Milky Way ISM parameters (OC97). In this 
case the mean number of supernovae per bubble is $ \simeq 9 $
and $m_* \simeq 1350\ M_\odot$. (We note that if $N_{\rm max}$ was an order
of magnitude greater than this, $m_*$ would only increase by $25 \%$).  
Taking $E_{\rm sn} \simeq 10^{51} $ ergs, we obtain:

\begin{equation}\label{qss3}
 Q_{\rm ss} \simeq \frac { 7 f_d\  {\rm SFR_\odot} }{M_{\rm ISM,{10}}\ \tilde v_{10}^2}
         \quad ,
\end{equation}

\noindent 
where $\rm SFR_\odot$ is the star formation rate in solar masses per year,
$M_{\rm ISM,{10}}$ is the mass of the ISM in units of $10^{10} M_\odot$ 
and $\tilde v_{10}$ is the thermal velocity of the ISM normalised to
$10$ km s$^{-1}$. 
In a system where star formation has been ongoing for a
time $t < t_e$, the porosity is given by (equation~\ref{vtt2}):

\begin{equation}\label{qt}
Q(t) = Q_{\rm ss} \Bigl(\frac{t}{t_e}\Bigr)^2
         \quad ,
\end{equation}

Equation~\ref{qss3} implies that there is a critical star formation
rate, $\SFR_{\rm crit}$ such that the porosity
of the ISM is unity, i.e.

\begin{equation}\label{SFRc}
 \SFR_{\rm crit} = 0.15 \biggl(\frac{ M_{\rm ISM,{10}}\ \tilde v_{10}^2}{f_d}\biggr)\ M_\odot {\rm yr}^{-1}
         \quad ,
\end{equation}

We stress that {\it $\SFR_{\rm crit}$ is the SFR such that the
energy output  from SNe, over a timescale $t_e$, is comparable with the
energy of the ISM contained in random motions.} The normalisation of 
equation~\ref{SFRc}  thus depends only on the assumed IMF and the stellar
astrophysics contained in the value of $t_e$ and the energy delivered
per SN. (We note, however, that in reality equation~\ref{SFRc} should
be regarded as a very rough guide, since its derivation suffers from
the obvious over-simplification that results from approximating
the ISM of a galaxy as a smooth homogeneous entity characterised by
a single set of physical parameters. In practice, we will find
equation~\ref{SFRc} useful below in dividing highly porous regimes
from the marginal case and from situations where the porosity
is very low).

If $\SFR < \SFR_{\rm crit}$, then such a SFR can be sustained
indefinitely, provided the gas supply is large; star formation can
proceed at such a rate over timescales 
$\gg t_e$, with the porosity attaining a steady state value of $Q_e < 1$.

  If $\SFR > \SFR_{\rm crit}$, then the system attains unit porosity after a
time $t_Q$:

\begin{equation}\label{tq}
t_Q = t_e \biggl({\SFR_{\rm crit}\over{\SFR}}\biggr)^{1/2} \quad .
\end{equation}
We discuss below the consequences of achieving unit porosity, but first
note that the {\it maximum} rate of star formation achievable in a
star forming system is
\begin{equation}\label{SFRd}
\SFR_{\rm dyn} \sim {{M_{\rm ISM}}\over{t_{\rm dyn}}} \quad ,
\end{equation}
where $t_{\rm dyn}$ is the dynamical timescale of the star forming region.

\section {Consequences for star formation efficiency and escape of ionising radiation}

\subsection{Star formation efficiency}

We have shown that the porosity of a star
forming system becomes of order unity at the point that the input of
mechanical energy into the ISM (over time $t_e$, or the duration of the
burst, whichever is the shorter)
is comparable with the {\it thermal energy content} of the ISM,
where we take `thermal' to denote random ISM motions.
The critical SFR that must be sustained over a timescale
$t_e$ in order to attain unit porosity is given by $\SFR_{\rm crit}$
(equation~\ref{SFRc}). 

  For a bound spheroidal system, the thermal energy content of the ISM
is always of order its gravitational binding energy, whether the
gravitational potential derives from the gas itself or is a
background potential of dark matter and/or stars. 
Consequently, when the
porosity attains a value $\sim 1$, the energy input into the ISM
is comparable with its gravitational binding energy. As a result, one
would not expect spheroidal systems to be able to sustain star formation
rates much in excess of $\SFR_{\rm crit}$ over timescales $> t_e$.

 In the case of compact systems with dynamical timescale $<t_e$, it is however
possible for the SFR to exceed this limit temporarily. If,
in this case, we consider the star formation event to be
self-terminated after time $t_Q$ (equation~\ref{tq}),
when the volume filling factor of superbubbles reaches unity, we can
use equations (19) and (23) to derive the  fraction of gas converted into stars during the event as: 
 \begin{equation}\label{eps1}
\epsilon = {{2 m_* \tilde v^2}\over{9 E_{\rm sn}}} 
        \biggl({{\SFR}\over{\SFR_{\rm crit}}}\biggr)^{1/2} \quad .
\end{equation}
so that
 \begin{equation}\label{eps}
\epsilon = 6 \times 10^{-4} \tilde v_{10}^2 
	\biggl({{\SFR}\over{\SFR_{\rm crit}}}\biggr)^{1/2} \quad .
\end{equation}
The maximum fraction of the ISM that can be turned into stars increases
with the square root of the SFR, and is thus limited by
the upper, dynamical, limit to the SFR implied by
equation~\ref{SFRd}, to the value:

\begin{equation}\label{epsmax1}
\epsilon_{\rm max} =   \biggl({{2 m_* \tilde v^2}\over{9 E_{\rm sn}}}\biggr)^{1/2}
        \biggl({{t_{\rm dyn}}\over{t_e}}\biggr)^{-1/2} \quad .
\end{equation}

\noindent from which

\begin{equation}\label{epsmax}
\epsilon_{\rm max} = 0.02  \tilde v_{10}  
	\biggl({{t_{\rm dyn}}\over{t_e}}\biggr)^{-1/2} \quad .
\end{equation}
 
  In a disc system, by contrast, the thermal energy content of the
ISM is  $\ll$ its gravitational binding energy.  Consideration of hydrostatic
equilibrium normal to the disc plane demonstrates that the ratio of
energy in random motions to gravitational binding energy is of
order  $(H/R)$  if the vertical gravity mainly derives from the
disc's local self-gravity, or $(H/R)^2$ if it instead derives from
the vertical component of the gravity of a central mass concentration.
Thus, star formation rates in excess of $\SFR_{\rm crit}$ do not
imply the wholesale unbinding of the ISM and may not be ruled out on
these grounds.

  Whether or not $SFR_{crit}$ represents a maximum to the star formation
rate in discs systems, or else a point of transition to star formation
in a highly porous state, depends on conditions in the cool gas component
once $Q \sim 1$ (i.e., equivalently, whether feedback operates positively
or negatively on the cool gas). Several pieces of evidence suggest that
star formation may well continue in this state. We will see below
(Section 4.3) that some systems, notably Lyman Break Galaxies, 
apparently sustain star formation rates well in excess of $SFR_{\rm
crit}$ over prolonged periods 
($10^8-10^9$ years; Shapley et al 2001).  A more local example is provided 
by the LMC.  Although the porosity of this system is around unity (Oey
et al 2001), continuous vigorous star formation has been ongoing in
the LMC disk for at least $10^9$ year, and possibly up to 15 Gyr
(e.g., Smecker-Hane {\etal}2002; Dolphin 2000).  Inspection of the LMC reveals
how this situation is achieved: although the bulk of the system volume
is filled with hot gas, chiefly in the halo, the bulk of the {\it mass}
is in the cool component in the disc plane, where a high
star-formation rate is maintained.  This system shows little evidence
for negative feedback effects on star formation, but is well-known to
provide examples where star formation appears to be actively triggered
in the cool gas constituting bubble walls (e.g., Yamaguchi
{\etal}2001; Oey \& Smedley 1996; Oey \& Massey 1995; Walborn \&
Parker 1992).  The specific example of the supergiant shell LMC-4,
which is the largest and most studied case of LMC triggered star
formation (e.g., Dopita et al. 1985; Braun et al. 1997; Olsen et
al. 2001) clearly shows a massive ring of H I and star formation 1.4
kpc in diameter surrounding a large complex of young blue stars.
However, the morphology is clearly ring-like, rather than shell-like,
suggesting that mechanical feedback on these large scales does not
remove the majority of shell gas from the low galactic latitudes, and
does promote continued star formation.  
In what follows, therefore, we assume that in disc systems star
formation can in principle continue in the cool component at rates
in excess of $SFR_{\rm crit}$.

 We note that the characteristic gas exhaustion timescale for
a system in a marginally porous state is given by:
\begin{equation}\label{texh}
 t_{\rm exh} = \frac{M_{\rm ISM}}{\SFR_{\rm crit}} = 7 \times 10^{10} {{ f_d }\over{\tilde v_{10}^2}}\ {\rm years} \quad ,
\end{equation}
where the value of $t_{\rm exh}$ depends on the same assumptions about
stellar astrophysics and IMF as detailed for $\SFR_{\rm crit}$ (equation~\ref{SFRc}).

\subsection{Implications for the escape of ionising radiation}

We have here presented a model of the ISM structured by SN explosions 
located in spatially distributed clusters, a model that has been 
successfully tested in the case of nearby galaxies. Our discussion
of ionising photon escape from such a a medium  
is necessarily more speculative, pending detailed hydrodynamic and
photoionisation calculations.  Here we  explore
the consequences of such a model by the following crude
parameterisation of ionising photon escape:
we assume that when the porosity of the
ISM is $< 1$, no ionising photons can escape and that when the porosity
is high ($>1$) all ionising photons can escape. 

It is easy to see why
these assumptions are wrong in detail.  For example, pure photoionisation
codes of disc galaxies, i.e. calculations that assume a smoothly
stratified initially neutral medium, suggest escape fractions of a
few per cent even in the absence of mechanical energy input from
SNe. Likewise, it is well established that populous clusters
can create local chimneys in the ISM, thereby launching galactic
superwinds and it is reasonable to expect some photon leakage in
this case (see, however, Tenorio-Tagle et al 1999; Dove, Shull and
Ferrara 1999) even when the global star formation rate is $\ll \SFR_{\rm crit}$.
It is also unlikely that the escape fraction is as high as unity even when
the ISM is highly porous. Although most of the volume of the ISM is
cleared of neutral material in this case,  most of its {\it mass}
is contained in neutral bubble walls. We
here assume that a variety of hydrodynamical instabilities break up the bubble
walls once the bubbles start to overlap strongly, thus opening up lines of
sight through which ionising photons can escape the disc.
In order to
escape the galaxy, however, such photons also have to propagate through 
low density material in the halo without encountering significant opacity
from neutral hydrogen. Detailed hydrodynamic/radiative transfer calculations
are required, which model the input of ionising photons and mechanical
energy into the halo from spatially dispersed, non-coeval star formation
events in the disc, in order to assess whether the halo can be maintained
in a state of sufficient transparency.   

  Despite the above caveats, we argue that this simple prescription
captures an important dependence of the escape fraction of ionising radiation
on SFR. Systems maintaining a steady state star formation
rate on timescales greater than $t_e$ can
exist in two states:
if $\SFR<\SFR_{\rm crit}$ the escape fraction is low and star formation
can in principle proceed at such a rate until all the gas is exhausted.
On the other hand, disc systems which are re-supplied on a timescale
less than $t_{\rm exh}$ (equation~\ref{texh}) may  sustain SFRs
in excess of  $\SFR_{\rm crit}$, in which case the escape fraction would be high.

For star-forming systems in which the dynamical
timescale is less than  $t_e$, the SFR
can vary on a timescale less than $t_e$, offering the possibility that
the SFR may temporarily exceed $\SFR_{\rm crit}$. During such
a star formation episode, the porosity of the system rises, attaining
unity at time $t_Q$. We have argued that in spheroidal systems, star
formation is self-limited at this point. Ionising photons from the
population created prior to this can then escape the immediate
vicinity relatively easily; the above crude model posits escape with
unit probability.  Given an IMF and relationships 
between ionising luminosity, mass and lifetime, one may readily calculate
an upper limit to the number of ionising photons escaping the region.
Specifically, if the number of ionising photons emitted by the population
prior to time $t$ is ${\cal{ N}}_{\rm ion}(t)$, then this upper limit is
given by:
\begin{equation}\label{fesc}    
f_{\rm esc} = 1 - {{{\cal {N}}_{\rm ion}(t_Q)}\over{{\cal {N}}_{\rm ion}(\infty)}}
	\quad .
\end{equation} 
Note that, whereas escape fractions are conventionally defined in 
terms of rates, in the case of a finite episode, it makes sense to
define the escape fraction in terms of {\it numbers} of ionising photons.

\begin{figure}
\includegraphics[width=3.5truein,height=3.5truein]{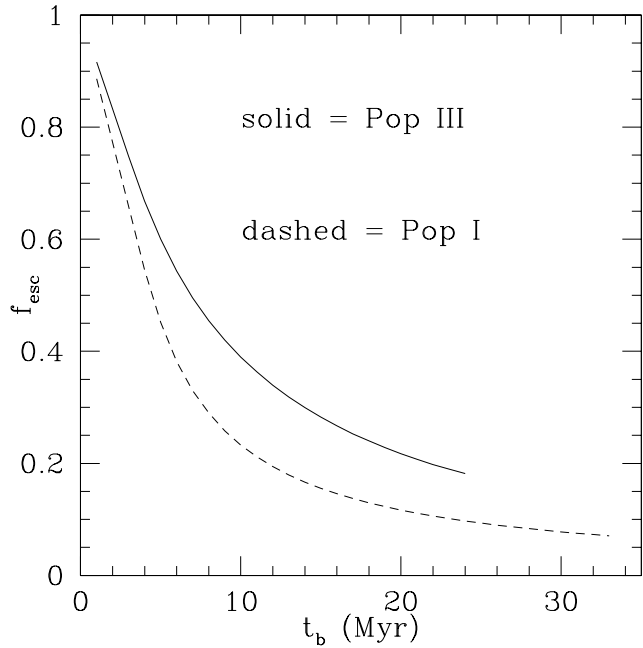}
\caption{\label{fesc}
Escape fraction (defined by equation~\ref{fesc} ) as a function of the time, $t_b$,
at which the burst of star formation is terminated. The solid and dashed curves
are for  Population III and   Population I stars, with a Salpeter IMF
(extending up to $100 M_\odot$) in both cases. 
}
\end{figure}

 Figure 1 illustrates the dependence of $f_{\rm esc}$, as defined above, on
the duration of a star formation burst ($t_b$) under the simple assumption
that ionising photons escape only if emitted
at times $> t_b$.
The two lines denote stars of Population I and III, with an
assumed Salpeter IMF in both cases.  Note that star formation
is assumed to continue at constant rate until $t_b$, unlike other
recent studies of escape of ionising radiation from
ageing populations (Dove, Shull and Ferrara 1999; Tenorio Tagle et al 1999)
where all the star formation is concentrated in a burst at time $t=0$. 
Stellar data for Population I stars is taken from Maeder 1990 and
D\'\i az-Miller et al 1998, whilst for Population III stars, 
models have kindly been supplied by Chris Tout in advance of
publication. The curve for Population II stars would be almost
 indistinguishable
from that for Population I, since the main dependence of ionising luminosity
on metallicity occurs for stars of sufficiently low mass that they make
a rather small contribution to the total ionising output of the cluster.
The Population III curve is rather different, however, since lower mass
stars are considerably hotter in this case and make a significantly larger
contribution to the total ionising output than for the Population I case.
Consequently,  a higher fraction of the ionising photons can escape at late
times in the Population III case. 
The effect, however, is not enormous: for bursts terminated after
$\sim 20$ Myr the escape fraction (equation \ref{fesc}) for Population III 
stars is $\sim 25 \%$
compared with a value that is roughly a factor of two lower for Population
I stars.

\section{Application to star forming systems}

\subsection{Compact systems} 

 The basic building block of star formation at the present epoch is the 
 Giant Molecular Cloud, and we here apply our simple model to assess the
internal self-destruction of GMCs through the action of SNe and
stellar winds (see Franco and Cox 1983 for an assessment of the effects
of winds from {\it low mass} stars on the dispersal of molecular clouds). For typical parameters
 ($M_{\rm ISM}=2 \times 10^5 M_\odot, \tilde v_{10} \sim 0.2$) 
it follows from equation~\ref{SFRc} that 
  $\SFR_{\rm crit}$ is a tiny  $ 10^{-7} M_\odot$ yr$^{-1}$, which is many orders
of magnitude less than the observed SFR in GMCs. The dynamical
timescale of GMCs is short (a few Myr; i.e $\sim 0.1 t_e$) so that if
the SFR is dynamically limited (equation~\ref{SFRd}), the maximum
fraction of the cloud that can be turned into stars is $\epsilon_{\rm max}
\sim 0.02$ (equation~\ref{epsmax}). This low value is similar to the fraction that
is generally inferred in molecular clouds (Larson 1988; Hunter 1992). 

 Following Whitworth (1979; see also Franco et al 1984, Yorke et al
1989), it is generally assumed that it is photoionisation  
that limits the efficiency of star formation in GMCs, prior to the
first SN explosions on a timescale of several Myr;
photoionisation is an effective agent of cloud disruption because of
the relatively shallow potential of GMCs.
Our estimate here suggests that, in solar metallicity
systems, the effect of clearing by stellar winds is likely to be
at least competitive with photoionisation during the 
first few Myr of cloud evolution. 
In practice,  simulations are required that include both  effects
since the trapping of ionisation fronts in bubble walls means that
photoionisation may be less effective than in the  homogeneous media
considered by previous authors 
(see discussion in D\'\i az-Miller et al 1999). 

 We may estimate the likely escape fraction from the immediate
vicinity of molecular clouds by combining the empirical lifetime
of molecular clouds ($\sim 10^7$ years; Leisawitz 1985; Fukui et al
1998) with Figure 1, from which
we deduce that only $10-20 \%$ of the Lyman photons
produced within a GMC will be emitted after the dispersal of the cloud.

 In low metallicity systems, however, the role of winds is much less
significant. Recent calculations of O star winds by Vink et al (2001)
indicate that the mechanical luminosity delivered by winds in stars
of given mass scales almost linearly with metallicity. Consequently,
although in Population I systems, winds deliver a mechanical luminosity
over the first few Myr that is comparable with the subsequent rate
of energy input from supernovae (Shull and Saken 1995; Leitherer and Heckman
1995), in low metallicity systems the input power from winds is negligible 
compared with supernovae. As a result, significant
feedback effects are delayed for
around $2$ Myr until the explosion of the first supernova. 
This factor
may explain the high inferred star formation efficiency in proto-globular 
clusters (Murray and Lin 1989; Geyer and Burkert 2001) and the compact
nature of globular clusters. More speculatively, we suggest that the
population of `faint fuzzies' (diffuse red clusters recently
discovered in early type galaxies; Larsen and 
Brodie 2000; Larsen et al 2001) may owe their relatively distended nature
and weak gravitational binding to the stronger winds that operate in clusters
whose metallicity is at the high end of the globular cluster metallicity
distribution. 

 In Population III systems, winds are estimated to be 
many orders of magnitude weaker than in Population I stars (Bromm et al
2001b), so that one may effectively ignore mechanical feedback from
winds in these systems. In the compact haloes that are expected
to host Population III stars, the dynamical timescale is sufficiently
short to allow efficient star formation prior to the explosion of
the first supernovae. Since Population III stars are likely to
be very massive 
(Bromm et al 1999, 2001a;
Abel et al 2000), the further evolution then depends on the details
of the resultant mass spectrum, since the energy output (and existence)
of supernovae in stars more massive than $100 M_\odot$ is highly
dependent on stellar mass (Bond et al 1984, Fryer et al 2001).
 
\subsection{Milky Way}

  For the ISM properties of the Milky Way ($n \sim 0.5$ cm$^{-3}$, $\tilde
v_{10} \sim 1$), the radius of a bubble stalling after $t_e$ is
$R_e \sim 1300$ pc (OC97), so that given the disc scale
height of $H \sim 100$ pc, the correction factor for disc systems
($f_d$; equation~\ref{fd}) is around $15 \%$. This means that the critical
SFR required to achieve unit porosity is boosted by
about $7$ relative to its value in an infinite medium, due to the loss
of accelerative power in bubbles that break out of the disc. For an ISM
mass of $\sim 10^{10} M_\odot$ in the Milky Way, the critical star formation
rate (equation~\ref{SFRc}) is roughly a solar mass per year,
that is, comparable
with the observed rate in the Milky Way (McKee and Williams 1997;
McKee 1989). As discussed above,
we may use our model as a crude indication of escape fraction:
`low' or `high' for SFRs that are much less than or
much greater than the critical value.   We would not however trust it in the
transitional case where the SFR is close to critical.
(See also the discussion in OC97 of the porosity of the Milky Way).

\subsection{Lyman Break Galaxies}

  The application of this model to Lyman Break Galaxies is currently
rather uncertain, given uncertainties about the gas masses and morphologies
of these objects. However, on the assumption that these are disc systems
($\tilde v_{10} \sim 1$) with gas masses comparable with their virial masses
($\sim 10^{10}\ \rm M_\odot$; Pettini et al 2001) one obtains values
of $\SFR_{\rm crit} \sim 1\ \rm M_\odot\ yr^{-1}$. The SFRs
inferred in Lyman Break Galaxies studied to date are comfortably greater
than this ($\sim 10-100\ \rm M_\odot\ yr^{-1}$; Pettini et al 2001) leading
to the expectation that ionising radiation should escape rather easily
from these systems. The discovery of Lyman continuum emission in the
composite spectra of Lyman Break Galaxies (Steidel et al 2001) is
consistent with this conclusion.  The positive correlation
between SFR and escape fraction that this model
predicts awaits the measurement of Lyman
continuum emission in individual Lyman break systems.

\subsection{Starbursts at low z}

  The nuclei of nearby starburst galaxies provide the best studied
examples of vigorous star formation activity in the local Universe;
the inferred SFRs  (up to
$\sim 20\ \rm M_\odot\ yr^{-1}\ kpc^{-2}$ for an assumed Salpeter
IMF; Lehnert and Heckman 1996; Meurer et al 1997)  are greatly in
excess of $\SFR_{\rm crit}$, leading to the 
expectation that the porosity of the ISM in these regions
should be high.  Note that this conclusion does {\it not}
depend on the IMF, but purely on the relationship between the
massive star content and ISM properties.  In the absence
of replenishment, therefore, one would expect a high fraction of
the ionising photons to be able to escape the nuclei of such
galaxies.

 The situation of starburst nuclei, located at the bottom  of the 
galactic potential well, however, means that such regions are likely
to be subject to continued replenishment of gas from larger radius
during the history of the starburst. The dynamical timescales in such
regions are short ($\sim 10^7$ years) and indeed less than or comparable
with $t_e$. Consequently, neutral material can flow into the region
on timescales less than $t_e$ and the filling factor
of regions devoid of \hi\ may not approach unity even at
the high star formation
rates typical of starbursts. Currently the available observational evidence
is that the escape fraction from starburst nuclei is indeed low  (Heckman et al
2001; Leitherer et al 1995). We highlight here the contrast with
the Lyman Break Galaxies (see above), where the more
extended star formation regions
do not permit re-supply on a timescale  of $t_e$.

\section{Conclusions}

  We have developed a model where the ISM porosity, i.e., the fractional
volume devoid of HI, is regulated by SN explosions. In this model,
the SN progenitors are located in spatially distributed OB
associations, membership numbers being dictated by the observed OB
association luminosity function.  This model has previously been shown to
provide a good fit to the observed size distribution of \hi\ holes in
nearby galaxies.

  We find that such a realistic distribution of SN progenitors
ensures that the clearing of the ISM is more effective, per unit star
formation rate, than in either the case of distributed single SNe
or the case where all SNe are concentrated in a single region. This
is because, given the slope of the OB association LF, the porosity
in our model is dominated by bubbles that come into pressure equilibrium
with the ISM on a timescale that is similar to $t_e$, where $t_e$ is the
maximum lifetime of a SN progenitor and hence the timescale
over which associations inject mechanical energy into the ISM. Such
superbubbles evolve quasi-adiabatically and thus most of the mechanical
energy of their SNe is deposited in the ISM. In consequence, for
the population of bubbles as a whole, the average volume cleared per
SN is within a factor of order unity of its theoretical maximum
(equation~\ref{vmax}), although this is somewhat reduced in the case of disc
galaxies (equation~\ref{fd}). This contrasts with the situation of single
SNe, where cooling limits the volume cleared, and also single burst
models, where clearing is limited by the breakout of the bubble from
the galactic plane. 

 This model yields a simple relationship between the star formation
rate and interstellar porosity. Following the arguments above, the critical
star formation rate ($\SFR_{\rm crit}$) required to attain a porosity 
of order unity is just that at which the energy input from SNe, over
a timescale $t_e\sim 40$ Myr, is comparable with the thermal energy content
of the ISM. For a given kinetic temperature of the ISM defined by
the level of random motions, this implies a simple relationship
between the $\SFR_{\rm crit}$ and the mass of the ISM
(equation~\ref{SFRc}), and hence a characteristic timescale for gas
exhaustion, $t_{\rm exh}$ (equation~\ref{texh}). 
For a Salpeter IMF and ISM velocity dispersion
of around $10$ km s$^{-1}$, this critical star formation timescale
is roughly $10^{10}$ years.

 If spheroidal galaxies form stars at $\SFR_{\rm crit}$, the energy input into
the ISM over $t_e$ is comparable with the gravitational binding energy of
the ISM and one might expect wholesale expulsion of the ISM to ensue. Disc
systems, by contrast, can remain in a highly porous state and still retain
their ISM.  Although in this case the volume fraction of \hi\ is then small,
the mass fraction is still large, so we assume that star formation can 
proceed in the shredded walls of interacting superbubbles.
 
  We furthermore suggest that the porosity of the ISM has an impact on
the escape of ionising photons from galaxies, since the disintegration
of overlapping bubbles
can create channels in the ISM through which ionising photons can escape.
This postulate must be assessed through detailed photoionisation
calculations in a medium structured by supernova explosions whose
progenitor OB associations are appropriately distributed in luminosity
and space.
We note that the recent analysis by Elmegreen et 
al (2001) of the morphology of the neutral ISM in the LMC favours 
the filament/bubble structure that is characteristic of a
supernova-structured ISM. 

 If we tentatively accept this postulate, we would thus expect high escape fractions in galaxies whose star formation
rates exceed $\SFR_{\rm crit}$, as would appear to be the case in Lyman Break
Galaxies. Sustained star formation at such rates however requires that
gas is replenished on a timescale less than $t_{\rm exh}$ (see above). At recent
cosmic epochs, the timescale for gaseous infall into galaxies is long,
so that one would not expect that galaxies in general should
display the high SFRs required to maintain a highly porous
ISM. This conclusion is consistent both with measurements of the 
HI hole size distributions in nearby galaxies and with the low 
leakage of ultraviolet photons from the Milky Way based on H$\alpha$ 
measurements of the Magellanic Stream (Bland-Hawthorn and Maloney 1999)
During the assembly
of galaxies at high redshift, however, much shorter infall timescales
are expected,  and we suggest that it is the continued infall of
material into Lyman Break Galaxies that allows them to sustain
vigorous star formation levels with a high associated escape fraction.
In systems where the infall timescale falls to values less than $t_e$,
however, the situation reverses, since the continual replenishment of
neutral material into  the star forming region can prevent the porosity from
ever attaining high values. We suggest that this is why local starburst
nuclei, being compact regions at the bottom of the galactic potential 
and thus subject to gaseous inflows on a short timescale, have a low escape
fraction despite their high rates of star formation per unit gas mass.  

  We also consider compact systems, with dynamical timescales
$< t_e$, in which the SFR may temporarily exceed
$\SFR_{\rm crit}$ and we have estimated the maximum number of Lyman continuum
photons that might be expected to escape these systems. The key factor
here is the efficacy of feedback from stellar winds prior to the
explosion of the first SN, which depends critically on metallicity.
In Giant Molecular Clouds with near solar metallicity, we find that winds
provide a very efficient means of cloud dispersal; we estimate that the
maximum fraction of the cloud mass that can be converted into stars prior
to their dispersal is a few per cent, close to the observationally inferred
value. This suggests that stellar winds may be at least as important
as photoionisation as a negative feedback mechanism in Giant Molecular Clouds.
In Population III systems, by contrast, the mechanical feedback from stellar
winds is negligible and clearing of the ISM is delayed until the
explosion of the first SN.  Supernovae are not expected for progenitors more
massive than $250 M_\odot$ however, so that an extremely top heavy IMF 
might result in inefficient clearing and a low escape fraction of
ionising radiation.

 Finally, we present these  calculations as a first
attempt to parameterise the relationship between escape fraction and SFR
in star-forming systems and suggest the utility of such
a prescription in semi-empirical models of galaxy formation and evolution.  
The critical star formation rate may also offer a useful means to
parameterize mechanical and chemical feedback.

\section{Acknowledgments}
We thank the referee, Andrea Ferrara, for comments that  improved
the paper, as well as 
Volker Bromm and
Max Pettini for their useful input. We are indebted to Chris Tout for providing 
Pop. III stellar models
prior to publication. 
MSO thanks the IoA, Cambridge for hospitality during this work.

\label{lastpage}

\end{document}